\documentclass[useAMS,usenatbib]{mn2e}
\usepackage{graphicx,lscape,amssymb}
\usepackage{natbib}
\usepackage{epstopdf}
\usepackage{fancyhdr}
\usepackage{subfigure}
\usepackage{longtable}
\usepackage{rotating}
\usepackage{calc}
\usepackage{fancyhdr}
\usepackage{float}
\usepackage{enumerate}
\usepackage{sidecap}

\usepackage{amsmath}
\usepackage{subfigure}
\usepackage{booktabs, dcolumn}
\usepackage[dvipsnames]{xcolor}
\usepackage{hyperref}

\newcommand{\Teff}{$T_{\mathrm{eff}}$}

\newcommand{\totinit}{$12\,359\ $}
\newcommand{\fullcat}{$11\,407\ $}

\title{A catalogue of white dwarf candidates in VST ATLAS}

\author[Gentile Fusillo et al.]{Nicola Pietro Gentile Fusillo$^1$, Roberto Raddi$^1$, Boris T. G\"ansicke$^1$, J.~J.~Hermes$^2$\thanks{Hubble Fellow},
\newauthor Anna F. Pala$^1$, Joshua T. Fuchs$^2$, Ben Chehade$^3$, Nigel Metcalfe$^3$,  Tom Shanks$^3$\\
$^1$ Department of Physics, University of Warwick, Coventry, CV4 7AL, UK\\$^2$ University of North Carolina, Chapel Hill, NC 27599-3255, USA\\$^3$ Department of Physics, Durham University, South Road, Durham DH1 3LE, UK}
\begin{document}
\maketitle

\label{firstpage}
\begin{abstract}
The Sloan Digital Sky Survey (SDSS) has created a knowledge gap between the Northern and the Southern hemispheres which is very marked for white dwarfs: only $\simeq 15\%$ of the known white dwarfs are south of the equator. Here we make use of the VST ATLAS survey, one of the first surveys obtaining deep, optical, multi-band photometry over a large area of the southern skies, to remedy this situation. Applying the colour and proper-motion  selection developed in \citet{gentilefusilloetal15-1} to the most recent internal data release (2016 April 25) of VST ATLAS we created a catalogue of $\simeq4200$ moderately bright ($g\leq19$), high-confidence southern white dwarf candidates, which can be followed up individually with both the large array of southern telescopes or in bulk with forthcoming multi-object spectrographs.  
\end{abstract}


\section{Introduction}
White dwarfs are the final stage of the evolution of stars with main sequence masses $M > 0.8M_\odot$ and $M\lesssim8-10 M_\odot$ \citep{ibenetal97-1}, a range which includes the vast majority of all stars.  White dwarfs are therefore key tracers of the evolutionary history of the Galaxy (e.g.\citealt{torresetal05-1, tremblayetal14-1}) and significant contributors to the global stellar population. However, to fully exploit the diagnostic potential of the Galactic white dwarf population, it is necessary to reliably constrain fundamental parameters such as their space density \citep{holbergetal02-1, holbergetal08-1, giammicheleetal12-1, sionetal14-1}, mass distribution \citep{bergeronetal92-1, liebertetal05-1, falconetal10-1, tremblayetal13-1, tremblayetal16-1} and luminosity function \citep{catalanetal08-2, giammicheleetal12-1, rebassa-mansergasetal15-1}. These studies require large, homogeneous, and well-defined samples which, given the intrinsic low luminosity of white dwarfs, are still challenging to be assembled.  

Large samples of white dwarfs are also the starting point in searches for rare sub-types like magnetic white dwarfs \citep{gaensickeetal02-5, schmidtetal03-1, kuelebietal09-1,kepleretal13-1, hollandsetal15-1}, pulsating white dwarfs (\citealt{castanheiraetal04-1, greissetal14-1, gentilefusilloetal16-1}, see Sect.\,\ref{pulse}), high/low mass white dwarfs \citep{vennes+kawka08-1, brownetal10-1, hermesetal14-1}, white dwarfs with unresolved low mass companions \citep{farihietal05-1, girvenetal11-1, steeleetal13-1}, white dwarfs with rare atmospheric composition \citep{schmidtetal99-1, dufouretal10-1,gaensickeetal10-1, kepler+koester16-1}, close white dwarf binaries \citep{marshetal04-1, parsonsetal11-1}, metal polluted white dwarfs \citep{sionetal90-1, zuckermanetal98-1, dufouretal07-2, koesteretal14-1, raddietal15-1} or white dwarfs with dusty or gaseous planetary debris discs \citep{gaensickeetal06-3,farihietal09-1, debesetal11-2, wilsonetal14-1,manseretal16-1}.

\begin{figure}
\centering
\includegraphics[width=\columnwidth]{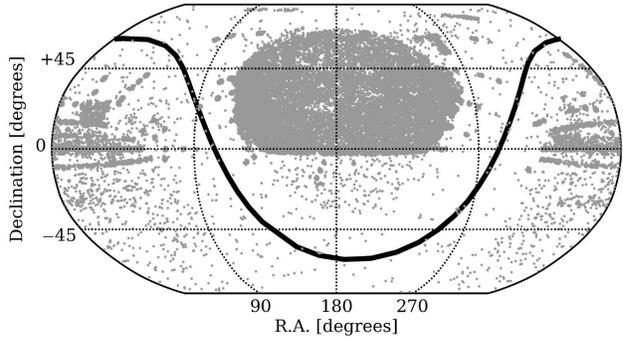}
\caption{\label{north_vs_south} Sky distribution of the $\simeq 39\,000$ white dwarfs confirmed to date. Only $\sim 15$\,per cent of them are located below the equator.}
\end{figure}

In recent years the number of known white dwarfs has increased by an order of magnitude, in particular thanks to the Sloan Digital Sky Survey (SDSS, \citealt{yorketal00-1}) which led to the identification of over 26\,000 white dwarfs mainly in the Northern hemisphere \citep{harrisetal03-1,  eisensteinetal06-1, Kleinmanetal13-1, kepleretal15-1, gentilefusilloetal15-1, kepleretal16-1}. The Southern hemisphere (below Dec $\simeq -20^\mathrm{\circ}$)  has not yet been surveyed by deep multi-colour CCD photometric surveys,  and consequently  only $\approx 15$ percent of all known white dwarfs are south of the celestial equator (cf Fig.\,\ref{north_vs_south}). However, the potential for identifying large numbers of white dwarfs in the Southern hemisphere is now rapidly growing thanks to the public surveys carried out by the European Southern Observatory (ESO) with the VLT Survey Telescope \citep[VST;][]{vst12-1}: ATLAS \citep[][]{shanksetal15-1}, VPHAS+ \citep{vphas+14-1}, and KIDS \citep{kids13-1}. In a pilot study we have  have identified white dwarfs at low Galactic latitudes by applying traditional color-cuts to VPHAS+ photometry \citep{Raddi16}. 
Here we present a catalogue of \fullcat colour-selected sources from ATLAS for which we calculated \emph{probabilities of being white dwarfs} ($P_\mathrm {WD}$) according to the method described in \citet{gentilefusilloetal15-1}. The $P_\mathrm {WD}$ values allow for selection of ATLAS white dwarf candidates with flexible efficiency and completeness, from which we estimate the catalogue to contain $\approx 4100$ genuine white dwarfs (Table\,\ref{summary_ATLAS}). 

In the following two sections, we briefly summarize the ATLAS survey and describe the properties of the photometric system, and how it compares to SDSS photometry. In Section\,\ref{ppm} we briefly outline the methodology used to combine photometry and proper motions to calculate $P_\mathrm {WD}$ values.
The catalogue of white dwarf candidates is presented in Section\,\ref{ATLAS_cand}. The completeness of the catalogue and the spectroscopic confirmation of some white dwarf candidates are discussed in Section\,\ref{diagnostics}. The last section is dedicated to our conclusions.

\begin{table}
\begin{center}
\caption{\label{summary_ATLAS} Summary of the white dwarf candidate selection in ATLAS.}
\begin{tabular}{lr}
\hline
\hline\\[-1.5ex]
ATLAS objects in initial colour cut & \totinit\\
\hspace{0.5cm} Of which with no proper motion & 952\\ 
Magnitude limit of final sample & $g\leq 19$\\
Final sample of white dwarf candidates (Sect.\,\ref{ATLAS_cand})& \fullcat\\
High confidence white dwarf candidates ($P_{\mathrm{WD}}\geq0.41$)& $\simeq4200$\\

Also in \citet{gentilefusilloetal15-1} catalogue & \\
of SDSS white dwarf candidates & 879\\
\hspace{0.5cm} Of which confirmed white dwarfs & 130\\
\hspace{0.5cm} Of which confirmed contaminants & 171\\
\hline
\hline

\end{tabular}
\end{center}
\end{table}

\section{VST ATLAS}
\begin{figure}
\centering
\includegraphics[width=\columnwidth]{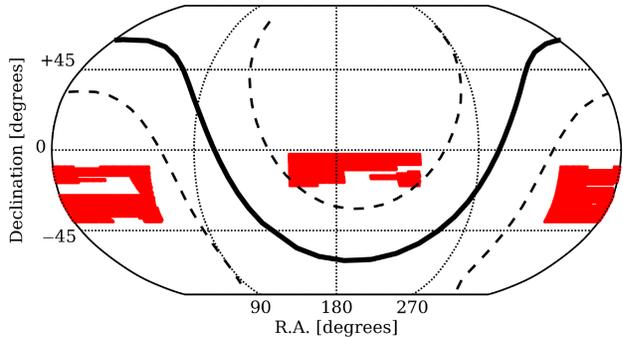}
\caption{\label{ATLAS_sky} Footprint of the sky area covered in all five filters ($ugriz$) by ATLAS at the time of the internal data release of April 25 2016. The solid black line indicates the location of the galactic plane and the dashed lines indicate regions $\pm 30^{\circ}$ from it.}
\end{figure}

VST ATLAS is primarily a cosmology-focused survey, aiming to image $4700\,\deg^2$ of the Southern Sky at high galactic latitudes ($|\mathrm{b}|>30^{\circ}$) in five bands ($ugriz$) to comparable depths to the SDSS in the North. The ATLAS footprint is divided into two contiguous blocks in the North and South galactic caps. The ATLAS South Galactic Cap (SGC) area lies between $21^h30^m<\,\mathrm{RA}<\,04^h00^m$ and $-40^{\circ}<\,\mathrm{Dec}<\,-10^{\circ}$, whilst the North Galactic Cap (NGC) area lies between $10^h00^m<\,\mathrm{RA}<\,15^h30^m$ and $-20^{\circ}<\,\mathrm{Dec}<\,−2.5^{\circ}$ plus $10^h00^m<\,\mathrm{RA}<\,15^h00^m$ and $-30^{\circ}<\,\mathrm{Dec}<\,-20^{\circ}$ (Fig. \ref{ATLAS_sky}).

The survey is carried out at the 2.6\,m VST, located at Cerro Paranal in Chile. The telescope mounts at the prime focus a 1\,sq-deg wide imaging instrument, the OmegaCAM \citep{Kuijken11}, which consists of 32 CCDs of $4{\rm k}\times2{\rm k}$ pixels each. The narrow gaps between the individual CCDs allow for an overall geometric filling factor of  91.4 per cent \citep[see][for more details]{shanksetal15-1}. The ATLAS band-passes are similar to those of the SDSS filters. Observations are taken in pairs for each filter and exposure times of 60\,s for $u$, 50\,s for $g$ and 45\,s for $ r, i$ and $z$. 
The imaging data is reduced by the Cambridge Astronomical Survey Unit (CASU) using the VST data flow software. Images are trimmed and debiased using nightly calibration frames and then flat-fielded using accumulated monthly stacked twilight sky flats. The frames are then corrected for cross-talk and defringed if necessary. The resulting imaging data comprise the combination of the two individual images for each of the original CCDs \citep{shanksetal15-1}.  For the analysis presented here,  we used the latest internal data release available on 2016 April 25. This release includes coverage in all five filters and photometric quality flags for $\simeq2400 \,$sq-deg of the sky, surpassing the publicly available data release 3.

\section{ATLAS v.s. SDSS}
\begin{figure}
\centering
\includegraphics[width=\columnwidth]{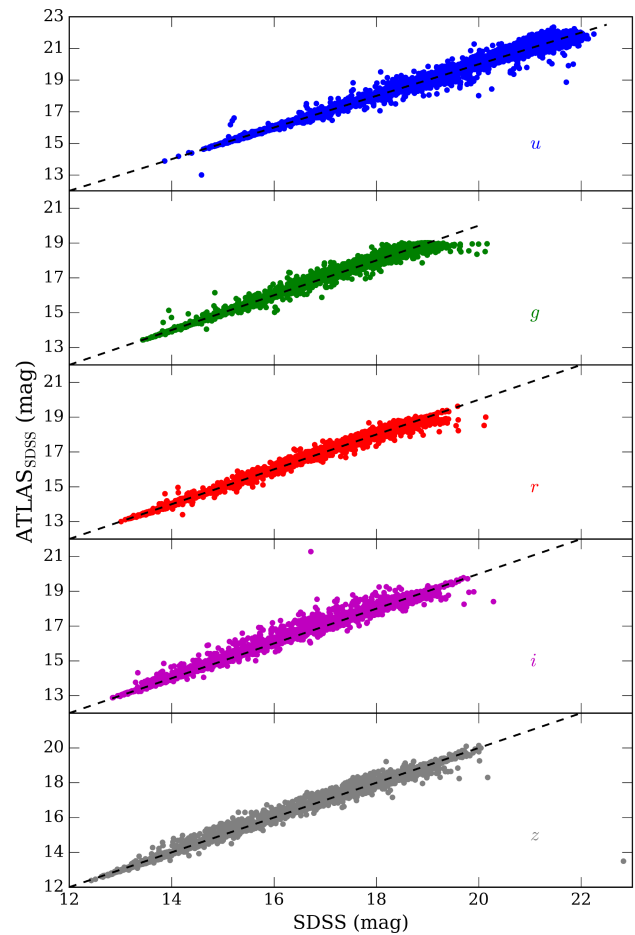}
\caption{\label{mag_comp_ATLAS} Comparison of ATLAS$_{\mathrm{SDSS}}$ and SDSS magnitudes for a sample of $\simeq 112\,000$ point sources. The dashed black lines indicate a 1:1 relationship. The comparison objects were chosen as SDSS objects with clean photometry.} 
\end{figure}

VST ATLAS uses the same optical filters as SDSS ($ugriz$) and in many ways aims to be the Southern hemisphere counterpart of SDSS.  However, though the filter systems are nominally the same, the actual filter transmission curves have small differences, the  detectors are not the same, the observing conditions at the telescope sites are different, and the flux calibration is conducted in different ways. As a result ATLAS and SDSS  magnitudes, and therefore colours, are not perfectly equivalent. As part of their recalibration of ATLAS photometry to the AB system \citet{shanksetal15-1} carried out a detailed comparison of SDSS and ATLAS photometry. ATLAS and SDSS overlap over an equatorial region of $\simeq300$ deg$^2$ covering parts of both the north galactic cap ($10^{h}\lesssim$ RA  $\lesssim$ 15$^{h}$\,30$^{m}$; $-3.5^{\circ}\lesssim$ Dec $\lesssim -2^{\circ}$) and the south galactic cap (22$^{h}$\,40$^{m	}$ $\lesssim$ RA $\lesssim$ 3$^{h}$; $-11^{\circ}\lesssim$ Dec $\lesssim -9^{\circ}$).  \citet{shanksetal15-1} used the objects in the northern galactic cap overlapping region to develop a set of colour dependent equations to convert ATLAS (AB) magnitude in equivalent SDSS magnitudes:
\begin{equation}
\label{colour-convert}
\begin{split}
&u_{\mathrm{SDSS}}=u_{\mathrm{ATLAS}}+0.01\times(u-g)+0.27\\
&g_{\mathrm{SDSS}}=g_{\mathrm{ATLAS}}+0.05\times(g-r)-0.06\\
&r_{\mathrm{SDSS}}=r_{\mathrm{ATLAS}}+0.03\times(g-r)-0.035\\
&i_{\mathrm{SDSS}}=i_{\mathrm{ATLAS}}-0.025\\
&z_{\mathrm{SDSS}}=z_{\mathrm{ATLAS}}-0.04\times(i-z)+0.04\\
\end{split}
\end{equation}
Since our selection method for white dwarf candidates makes use of a probability map in reduced proper motion-colour space that was initially developed from SDSS data (see Sect.\,\ref{ATLAS_cand}), it is of paramount importance to have reliable SDSS-equivalent ATLAS magnitudes (ATLAS$_{\mathrm{SDSS}}$ from here on). In order to evaluate the robustness of the magnitude transformations developed by \citet{shanksetal15-1}, in particular their applicability to blue objects, we carried out some further comparison with SDSS. We retrieved the available SDSS photometry of all  ATLAS sources in the overlapping regions with clean $g \leq 19.5$ SDSS photometry ($\simeq 112\,000$ objects). We then applied equations\,\ref{colour-convert} to the ATLAS photometry and compared the ATLAS$_{\mathrm{SDSS}}$  magnitudes with the SDSS ones (Fig.\,\ref{mag_comp_ATLAS}). We find that the mean values of SDSS $-$ ATLAS$_{\mathrm{SDSS}}$ magnitudes for the objects in our overlapping samples are: $u=0.0109\pm0.0003$, $g=0.0089 \pm0.0001$, $r=0.0086\pm0.0001$, $i=0.0098\pm0.0002$, $z=0.011\pm0.0003$. These mean differences are smaller than the typical  uncertainties in the SDSS and ATLAS magnitude. We therefore conclude that ATLAS$_{\mathrm{SDSS}}$ magnitudes  are, for most intents and purposes, equivalent to SDSS ones and our selection method for white dwarf candidates \citep{gentilefusilloetal15-1} can be directly applied to them. 

\section{Colour selection and proper motions}
\label{ppm}

Using the free form SQL query tool available on the OmegaCAM Science Archive webpage we retrieved  photometry for all ATLAS sources which have been observed in all five filters,  marked as ``stellar" or  ``probable stellar" and with no ``important"\footnote{as defined on the quality bit flags description at http://osa.roe.ac.uk} quality issue (Table\,\ref{sql-cut}).
We then applied the magnitude conversions described by equations\,\ref{colour-convert} to calculate ATLAS$_{\mathrm{SDSS}}$ magnitudes for all our sources. 
The first step in our photometric selection method for white dwarf candidates involves applying a set of colour constraints which broadly select all blue sources (Table\,\ref{col-cut}). These colour-cuts are designed to include all white dwarfs with $T_{\mathrm{eff}}\gtrsim7000\,\mathrm{K}$ and are required to reduce the initial sample to a more manageable size, but they are not sufficient to eliminate contamination from QSO and other blue objects (i.e. subdwarfs, A stars; for more details see \citealt{gentilefusilloetal15-1}).
This initial broad colour selection resulted in a sample of \totinit blue ATLAS sources.
ATLAS does not provide proper-motion measurements thus we decided to retrieve those  from the recently published Absolute Proper motions Outside the Plane (APOP, \citealt{qietal15-1}) catalogue. APOP proper motions are calculated from carefully re-reduced photographic plates from the STScI Catalog of Objects and Measured Parameters from All-Sky Surveys (COMPASS) archive of the GSC-II project \citep{laskeretal98-1}. APOP covers 22\,525 square degrees and provides proper motions for 100\,774\,153 objects to the limiting magnitude of R $\simeq 20.8$ with typical uncertainties ranging between 4 and 9\,mas/yr.
However the astrometry of APOP and ATLAS correspond to observations taken several years apart and most white dwarfs have high proper motions, typically ranging from 20 mas/yr to 200 mas/yr. White dwarfs can therefore move significantly over a few years to decades and a simple cross match between ATLAS and APOP using a  fixed matching radius can easily lead to several mis-matches or  missing objects.

\begin{table*}
\caption{\label{sql-cut} SQL casjob flags used to select ATLAS point sources with reliable photometry from the OmegaCAM Science Archive webpage.}
\begin{tabular}{ll}
\hline
Constrain & Effect \\
\hline
(mergedClass =--1) OR (mergedClass =--2) & selects objects marked as \\
& ``stellar" or ``probable stellar"\\
AND uppErrBits $|$ gppErrBits $|$ rppErrBits $|$ ippErrBits $|$ zppErrBits) $<$ 65536 & exclude sources with any\\ & ``important" quality issues\\
 OR (uppErrBits $|$ gppErrBits $|$ rppErrBits $|$ ippErrBits $|$ zppErrBits) $\&$ 0x00400040 != 0 & does not exclude  ``source within a \\
 & dither offset of the stacked frame\\
 & boundary"\\
\hline
\end{tabular}
\end{table*}

\begin{table}
\centering
\caption{\label{col-cut} Equations describing the colour and magnitude 
constraints used to select sources in the ATLAS footprint. The colour cuts were applied to the ATLAS magnitudes after converting them into SDSS equivalent ones.}
\begin{tabular}{lcl}
\hline
Colour & &  constraint \\
\hline
$(u-g)$ & $\le$ & $3.917 \times (g-r) + 2.344$\\
$(u-g)$ & $\le$ & $0.098 \times (g-r) + 0.721$\\
$(u-g)$ & $\ge$ & $1.299 \times (g-r) - 0.079$\\
$(g-r)$ & $\le$ & $0.450$\\
$(g-r)$ & $\ge$ & $ 2.191 \times (r-i)-0.638$\\
$(r-i)$ & $\le$ & $-0.452 \times (i-z) + 0.282$\\
$g$     & $\le$ & 19 \\
\hline
\end{tabular}
\end{table}

\begin{figure}
\centering
\includegraphics[width=\columnwidth]{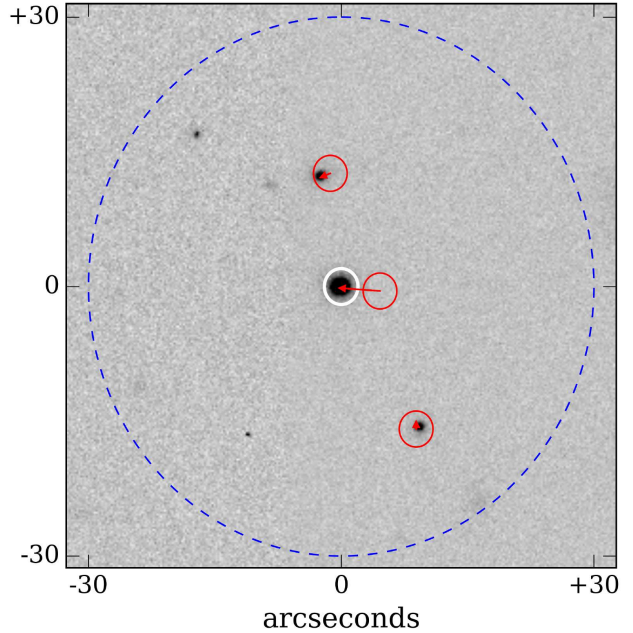}
\caption{\label{ATLASvsppmxl} ATLAS $g$ band image centred at the position of one of our white dwarf candidates. The blue circle represents the 30" radius area used for the first cross-match with APOP. 2" radius circles are shown centred on  the J2000 APOP coordinates of all matching sources in the initial cross-match and the red arrows indicate how the objects moved between J2000 and the ATLAS epoch of observation. The white circle indicates the final 2" matching radius around the ATLAS source. }
\end{figure}

We therefore divided our cross-matching procedure in three separate steps. For each ATLAS object we first retrieved every matching APOP source within a radius of 30 arcseconds (typically 4 to 8 objects) and compared the modified Julian date (MJD) of the ATLAS observation with that of APOP  (by definition at epoch J2000 so MJD 51544). We defined an epoch difference $\Delta t=\mathrm{MJD}_\mathrm{ATLAS} - 51544$ and then used the proper motions and J2000 positions from APOP to compute predicted positions at the epoch of the ATLAS imaging for all objects in the first cross-match (Fig.\,\ref{ATLASvsppmxl}). This coordinate ``forward projection'' is carried out according to:

\begin{equation}
\alpha=\alpha_{	\mathrm{APOP}}+\left(\frac{\mu_\alpha}{\cos(\delta_{\mathrm{APOP}})}\right)
\times
\frac{\Delta t}{365.25}
\end{equation}

\begin{equation}
\delta=\delta_{\mathrm{APOP}}+\mu_\delta \times\frac{\Delta t}{365.25}
\end{equation}													
where $\mu_\alpha$ and $\mu_\delta$ are the objects proper motions in right ascension and declination respectively. Finally we consider a true match  to be the closest object whose forward projected coordinates fall within two arcseconds of the ATLAS ones. In cases where more than one matching object is found within two arcseconds (a few tens within the entire sample) we select the best match by visually inspecting the magnitudes of the matching pairs  and their angular separation.

Following this procedure we obtained proper motions for \fullcat\ objects. The most likely explanation for the 952 ATLAS objects for which we could not find a counterpart in APOP is that they could not be reliably matched up on the photographic plates used by APOP. 

\section{White Dwarf candidates selection}
\label{ATLAS_cand}
In order to identify reliable white dwarf candidates among ATLAS sources  we rely on the photometric selection method presented in \citet{gentilefusilloetal15-1} which can be used to a assign a ``\emph{probability of being a white dwarf}" ($P_{\mathrm{WD}}$) to any object with available multi-band photometry and proper motion. 
In this section we briefly summarize the details of the selection method; for a full description refer to  \citet{gentilefusilloetal15-1}.
The $P_{\mathrm{WD}}$ values rely on a probability map which traces the distribution of spectroscopically confirmed white dwarfs and contaminant objects selected from SDSS in colour and reduced proper motion $H$ computed as:
\begin{equation} H_g=g+5\log\mu+5\end{equation}
where $\mu$ is the proper motion in arcsec/year. This probability map effectively traces which areas in colour-$H$ space are more likely populated by either white dwarfs or contaminants. In our work on SDSS photometry we determined that the strongest discrimination between white dwarfs and contaminants is obtained in the $g-z, H_g$ space, which we therefore adopted for our selection method. The final map  was constructed using a training sample of over 27\,000  objects (different types of white dwarfs, quasars and stellar contaminants) that were classified by visual inspection of their SDSS spectra.  By combining the $(g-z, H_g)$ position of a test object with this probability map we can compute a quantity that directly indicates how likely it is for the object to be a white dwarf, in other words our  $P_{\mathrm{WD}}$. 
We have shown above that ATLAS$_{\mathrm{SDSS}}$ magnitudes are equivalent to the SDSS ones. We therefore calculated $H_g$ for all ATLAS  objects using the ATLAS$_{\mathrm{SDSS}}$ magnitudes and the APOP proper motions and directly applied the \citet{gentilefusilloetal15-1} selection method to calculate $P_\mathrm {WD}$ for all \fullcat\ ATLAS sources in our sample. In Table\,\ref{Col_tab} we summarize the content of our final catalogue of ATLAS white dwarf candidates. 

We also performed a cross-match of our catalogue with the Gaia DR1 source catalogue \citep{gaiaDR1_16} and provide Gaia source ID and $G$-band mean magnitude for all matching sources. Gaia is able to resolve objects with a sky separation of 0.23" \citep{debruijneetal15-1}, a resolution much higher than what achievable by VST ground based observations. As a result we found two ATLAS objects (ATLASJ235435.65$-$290704.08 and ATLASJ121100.93$-$075241.23) which were each matched to two Gaia sources both with an angular separation of $<1"$. These objects are likely to be binary systems which were resolved with Gaia, but not in ATLAS. ATLASJ121100.93$-$075241.23 could be of particular interest being a relatively bright white dwarf candidate ($P_\mathrm {WD} = 0.71$, $g = 15.9$) with a potential faint close companion ($G=18.4$). 
Out of the five ATLAS bands we find that $r$  is the one closest to Gaia $G$ particularly for sources with $g-r \geq 0$ where the mean difference $G$-$r$  is $0.12$ mag.  

\begin{table*}
\centering
\caption{\label{Col_tab} Format of the  catalogue of VST ATLAS white dwarfs candidates. The full catalogue can be accessed online via VizieR. }
\begin{tabular}{lll}
\hline
\hline
Column No. & Heading & Description\\
\hline
1 & VST ATLAS name & ATLAS objects name (ATLAS + J2000 coordinates)\\
2 & ATLAS ID & Unique ID identifying the photometric source in ATLAS\\
3 & ra & right ascension\\
4 & dec & declination\\
5 & $P_\mathrm {WD}$  & The \emph{probability of being a WD} computed for this object\\
6 & umag & ATLAS $u$ band magnitude\\
7 & umag err & ATLAS $u$ band magnitude uncertainty\\
8 & gmag & ATLAS $g$ band magnitude\\
9 & gmag err & ATLAS $g$ band magnitude uncertainty\\
10 & rmag & ATLAS $r$ band magnitude\\
11 & rmag err & ATLAS $r$ band magnitude uncertainty\\
12 & imag & ATLAS $i$ band magnitude\\
13 & imag err & ATLAS $i$ band magnitude uncertainty\\
14 & zmag & ATLAS $z$ band magnitude\\
15 & zmag err & ATLAS $z$ band magnitude uncertainty\\
16 & MJD & Modified julian date of ATLAS observation\\
17 & pmra & APOP proper motion in right ascension (mas/yr)\\
18 & pmra err & APOP proper motion in right ascension uncertainty (mas/yr)\\
19 & pmdec & APOP proper motion in declination (mas/yr)\\
20 & pmdec err & APOP proper motion in declination uncertainty (mas/yr)\\
21 & human class & classification of the object based on  inspection of its available spectrum (section\,\ref{spec_follow})\\
22 & Simbad type1 & currently available primary Simbad classifications\\
23 & Simbad type2 & currently available secondary Simbad classifications\\
24 & Gaia ID & Gaia DR1 source ID\\
25 & Gmag & Gaia DR1 $G$-band mean magnitude\\
\hline
\end{tabular}

\end{table*}

\section{Discussion} 
\label{diagnostics}
\subsection{Comparison with SDSS}
\label{SDSS_wd_comp}
In \citet{gentilefusilloetal15-1} we used an independent sample of spectroscopically confirmed white dwarfs and contaminants from SDSS DR9 and DR10 and later LAMOST DR3 \citep{gentilefusilloetal15-2} to demonstrate the efficiency of the selection method and the completeness of our catalogue of SDSS white dwarf candidates. However, similarly large spectroscopic samples do not exist for the southern hemisphere and therefore we cannot test in the same way the robustness of the selection method when applied to ATLAS photometry. Nonetheless, as a result of the overlap of ATLAS with SDSS,  879 objects appear in both the \citet{gentilefusilloetal15-1} catalogue of SDSS white dwarf candidates and in the ATLAS catalogue presented here. This sample includes 130  white dwarfs and 171 contaminants confirmed by SDSS spectroscopy (as of SDSS DR12 \citealt{alametal15-1}) which enable us to carry out some valuable tests on the ATLAS sample of white dwarf candidates.

\begin{figure}
\centering
\includegraphics[width=\columnwidth]{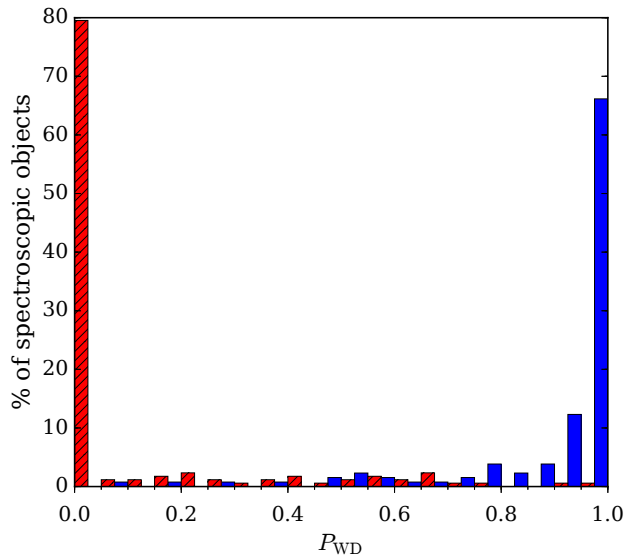}
\caption{\label{histo_ATLAS} Distribution of 301 spectroscopically confirmed white dwarfs (blue) and contaminants (red, shaded) from the SDSS and ATLAS overlap sample as a function of $P_{\mathrm{WD}}$.}
\end{figure}

Figure\,\ref{histo_ATLAS} shows that the vast majority of the 130 white dwarfs have $P_{\mathrm{WD}}$ (ATLAS) $>0.8$ while over 85 per cent of the 171 contaminants have $P_{\mathrm{WD}}$ (ATLAS) $<0.2$. Though this test is limited to small sample sizes, it is evident that the $P_{\mathrm{WD}}$ calculated from ATLAS and APOP data provide a clear discrimination between white dwarfs and contaminants. 

Using the same spectroscopic sample we can also calculate that a confidence cut which includes all ATLAS objects with $P_{\mathrm{WD}} \geq 0.41$  results in a 96 per cent completeness and 87 per cent efficiency in selecting white dwarfs. These numbers are very similar to those obtained from the catalogue of SDSS white dwarf candidates \citep{gentilefusilloetal15-1} when applying the same cut in $P_\mathrm{WD}$. We also compared the surface density of ATLAS and SDSS white dwarf candidates with $P_{\mathrm{WD}} \geq 0.41$ and for both samples we  find an average of $\simeq 1.8$ objects  per square degree. These results suggests that our catalogue of ATLAS white dwarf candidates should be as complete and reliable as the SDSS catalogue presented in \citet{gentilefusilloetal15-1}.

\begin{figure*}
\includegraphics[width=2\columnwidth]{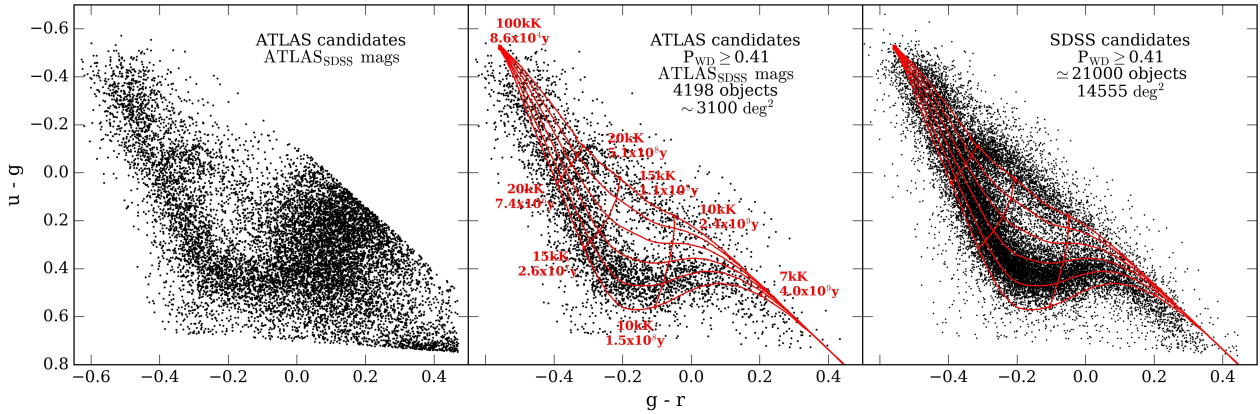}
\caption{\label{ATLAS_res} $u-g, g-r$ colour-colour distribution of (from left to right): all \fullcat  ATLAS objects in our final sample; 4205 ATLAS objects with $P_{\mathrm{WD}} \geq 0.41$; $\simeq21\,000$  white dwarf candidates from \citet{gentilefusilloetal15-1} with $P_{\mathrm{WD}} \geq 0.41$ for comparison. White dwarf cooling tracks from \citet{holberg+bergeron06-1} are shown in red overlay.}
\end{figure*}

The common ATLAS and SDSS white dwarf candidates  also allow us to directly compare $P_{\mathrm{WD}}$ values calculated using ATLAS and APOP with those calculated using SDSS data. We find that the $P_{\mathrm{WD}}$ values are largely consistent with an average difference $|P_\mathrm{WD}\mathrm{(ATLAS)}-P_\mathrm{WD}\mathrm{(SDSS)}|=0.042\pm 0.03$. However, $\simeq 4$ per cent of the objects in the overlapping SDSS and ATLAS sample show significantly inconsistent $P_\mathrm{WD}$ values,  $|P_\mathrm{WD}\mathrm{(ATLAS)}-P_\mathrm{WD}\mathrm{(SDSS)}| \geq 0.2$. Close inspection of these objects reveals that the cause of such difference in $P_{\mathrm{WD}}$ is a marked discrepancy in the SDSS and APOP proper motions, potentially caused by erroneous matching on the original photographic plates used by the surveys. Additionally, despite our best efforts we cannot fully exclude that a limited number of ATLAS objects may have been matched to the wrong APOP object (see Sec.\,\ref{ppm}) leading to a wrong assumed proper motion.
Even accounting for this small number of inconsistencies, we are confident that the $P_{\mathrm{WD}}$ values calculated can be used to reliably select high-confidence dwarf candidates, i.e. Fig.\,\ref{ATLAS_res} clearly illustrate that the colour-colour distribution of the ATLAS $P_\mathrm{WD} \geq 0.41$ sample is remarkably similar to that of the equivalent sample selected from the \citet{gentilefusilloetal15-1} SDSS catalogue. Taking into account the values of completeness and efficiency calculated before, we estimate that our catalogue contains $\simeq 4100$ high-confidence white dwarf candidates.

\subsection{Spectroscopic follow-up}
\label{spec_follow}
To further test the reliability of our selection method we obtained spectra for a total of 185 objects from our catalogue. 169 objects were observed with the Two Degree Field (``2dF") multi-object system of the AAOmega spectrograph on the Anglo Australia Telescope (AAT). These spectra were acquired as part of the 2dF Quasar Dark Energy Survey pilot \citep[2QDESp;][]{chehadeetal16-1}. The observations were made using the 580V and 385R gratings for the blue and red arm of the spectrograph respectively. This configuration achieves a useful wavelength range between 3700 and 8800 \AA . The data reduction was carried out using the 2dFDR\footnote{http://www.aao.gov.au/science/software/2dfdr} data reduction pipeline (for more details see \citealt{chehadeetal16-1}). Among these  169 targets we identified 14 new white dwarfs, all of which have $P_{\mathrm{WD}} >0.7$. The remaining objects are mostly quasars with $P_{\mathrm{WD}} <0.2$ and only four of them have $P_{\mathrm{WD}} >0.45$.

We also selected 16 additional targets specifically  as high-confidence white dwarf candidates ($P_{\mathrm{WD}} \geq 0.85$) and observed them with the NTT and the VLT as part of backup programs due to a northern pointing restriction for strong northerly winds. 13 targets were observed on 2015 September 16 using the EFOSC2 instrument on the NTT at la Silla, Chile with the 'Gr\#7' grism and a 1-arcsec slit, and with exposure times in the range of 300--900\,s. We carried out optimal spectral reduction and calibration using the packages PAMELA \footnote{PAMELA was written by T. R. Marsh and can be found in the STARLINK distribution Hawaiki and later releases.} and MOLLY \footnote{MOLLY was written by T. R. Marsh and is available from http://www.warwick.ac.uk/go/trmarsh/software.} \citep{marsh89-1}. The last 3 objects where observed on 2015 September 24 at the VLT observatory with the X-Shooter spectrograph, using a 1-arcsec slit  for the UVB arm and 0.9-arcsec for the VIS arm and exposure times of $\sim1500$\,s. The spectra were reduced using the standard procedures within the REFLEX\footnote{http://www.eso.org/sci/software/reflex/} reduction tool developed by ESO.  All 16 high-confidence white dwarf candidates were confirmed as white dwarfs (Table\,\ref{spec_table}). Both the NTT and the VLT observations were undertaken as backup programs due to a northern pointing restriction for strong northerly winds.

\subsection{Spectral analysis}
\begin{figure*}
\includegraphics[width=1.7\columnwidth]{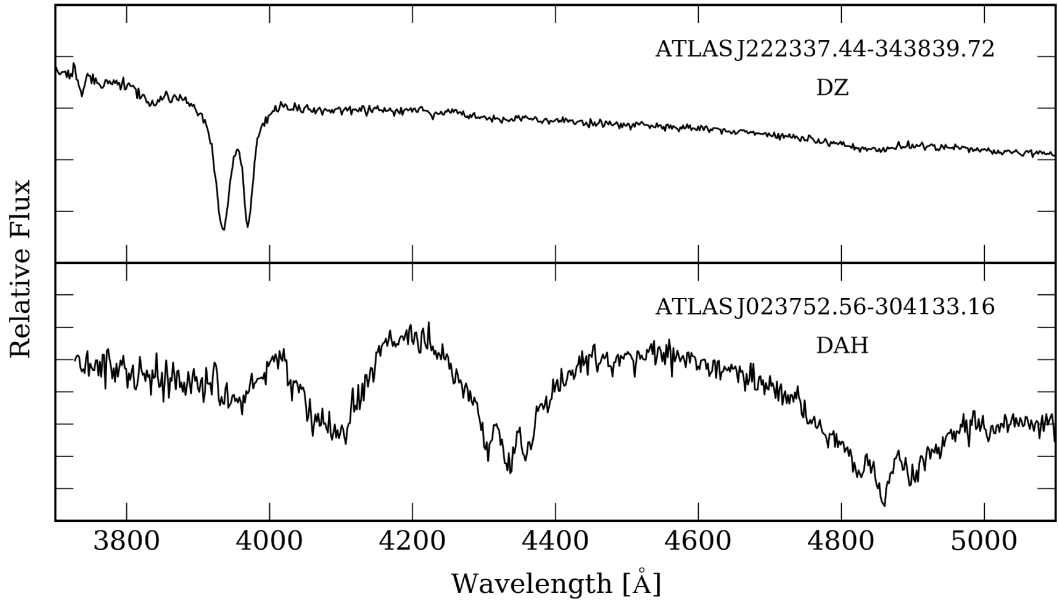}
\caption{\label{other_spec} Spectra of the DZ white dwarf and of one magnetic white dwarfs discovered by follow-up observations of candidates using EFOSC2.}
\end{figure*}

Of the 30 new spectroscopically confirmed white dwarfs, 27 stars have hydrogen dominated atmospheres (DA), one shows strong Ca H\&K lines (DZ, Fig\,\ref{other_spec}), one has a likely carbon-dominated atmosphere (DQ), and another star does not show strong atmospheric features at the signal-to-noise level of the spectrum we obtained. Two DA white dwarfs display also Zeeman splitting of the hydrogen lines due to moderately strong magnetic fields (DAH, e.g. Fig\,\ref{other_spec}).

\begin{figure*}
\includegraphics[width=2\columnwidth]{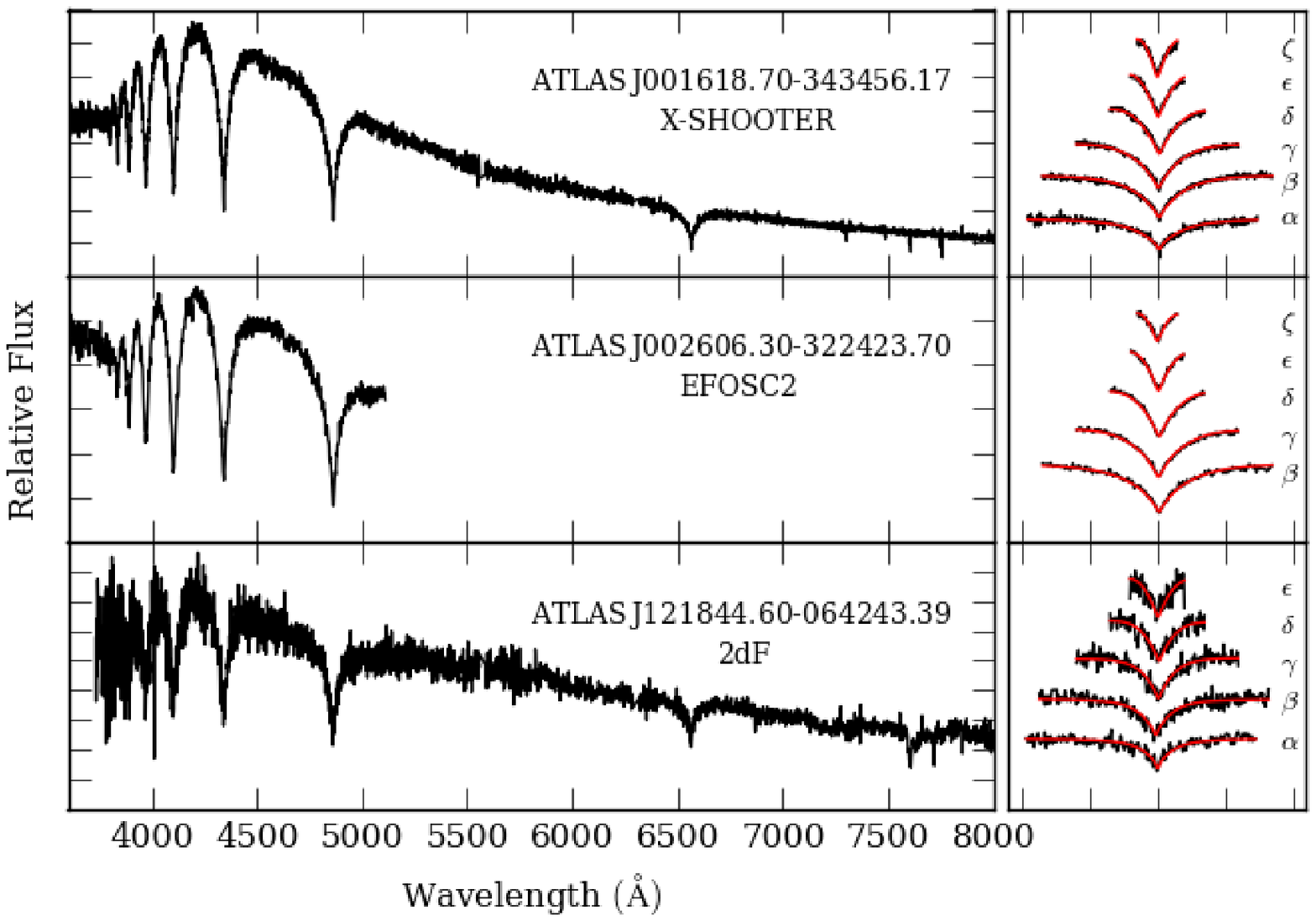}
\caption{\label{da_sample} Sample spectra of three white dwarf candidates confirmed by observations with X-SHOOTER, EFOSC2 and 2dF. The panels on the right show the best fitting models overlaid on the normalised Balmer lines used for the fit.}
\end{figure*}

In Table\,\ref{spec_table}, we summarize the spectral classification and we report the atmospheric parameters ($T_{\rm eff},\,\log{g}$) of the DA white dwarfs, which we have measured through comparison with a grid of \citet{koester10-1} model spectra (Fig.\,\ref{da_sample}). The synthetic spectra were computed with the mixing-length prescription of ${\rm ML2}/\alpha = 0.8$, and include the Stark broadening profiles by \citet{tremblay+bergeron09-1}. For the spectral analysis, we used {\sc fitsb2} \citep{Napiwotzki04} that determines the best-fitting model via $\chi^2$ minimisation of the Balmer line profiles for observed and synthetic spectra, using a downhill simplex algorithm \citep[e.g.\ the AMOEBA routine;][]{Press92} and a bootstrap method to assess the uncertainties. For cool DA white dwarfs (\Teff $<$ 15\,000\,K) we applied the \citet{tremblayetal13-1} 3D corrections of the atmospheric parameters to account for the inaccurate treatment of convention in 1D models.

The spectroscopic parameters are broadly consistent with the photometric estimates one would derive from comparison with the white dwarf cooling sequences (Fig.\,\ref{ATLAS_res}).

\subsection{New pulsating white dwarfs}
\label{pulse}
As it continues its tour around the ecliptic plane, the extended {\em Kepler} mission ({\em K2}) has opened the possibility to observe many new white dwarfs, especially those that pulsate. We have utilized this catalogue of candidate white dwarfs from ATLAS for target selection of several Guest Observer proposals (for Field 6, 12, and 15 in {\em K2} Campaign 6).
One of our candidates, selected solely based on its  $P_{\mathrm{WD}}$ and ATLAS $ugr$ colors, was observed to pulsate: ATLASJ134211.62$-$073540.1 (EPIC 229227292). In fact, this star became the fourth white dwarf to show aperiodic, large-amplitude outbursts in its {\em K2} observations \citep{belletal16-1}. Follow-up spectroscopy from the SOAR telescope confirmed this is a DA white dwarf with atmospheric parameters corresponding to 11\,190$\pm$170\,K, log g = 8.02$\pm$0.05, $M_{\mathrm{WD}}$ = 0.62$\pm$0.03. This is now the second-brightest white dwarf known to show such outbursts, which may arise  result from a parametric resonant coupling \citep{hermesetal15-1}.

Additionally, several of the white dwarfs analysed in Table\,\ref{spec_table} have temperatures and gravities near the empirical DAV instability strip. We followed-up four of these stars with high-speed photometry from the Souther Astrophysical Research Telescope (SOAR) at Cerro Pachon in Chile.  All targets were observed with the Goodman spectrograph in imaging mode using 20\,s exposures through an S8612 filter.
Three of the observed white dwarfs do not show photometric variability, with good limits on a lack of pulsations. ATLASJ023320.65$-$320310.88 was observed for 2.0\,hr and does 
not vary to a limit of 0.8\,ppt (1\,ppt = 0.1\,per cent). 
ATLASJ214039.37$-$341920.25 was observed for 2.4\,hr and does not vary to a limit of 2.0\,ppt. 
ATLASJ224510.44$-$383645.71 was observed for 2.1\,hr and does not vary to a limit of 2.9\,ppt.
However, we have detected significant variability in a 1.8\,hr run on ATLAS224653.56$-$385651.24: a 4.9(3)\,ppt peak at $1502.0\pm10.3$\,s. If confirmed, this would be one of the coolest (and longest-period) pulsating white dwarfs detected to date.
Within the uncertainties in  \Teff\ and log g (Table\,\ref{spec_table}) the two pulsating white dwarfs can be placed inside of the empirical ZZ Ceti instability strip and similarly the three stars observed not to vary can be placed outside it.

\begin{table*}
\caption{\label{spec_table} List of ATLAS white dwarf candidates confirmed by spectroscopic observations. For DA white dwarfs we also report the \Teff\ and $\log g$ from the best fitting model adjusted using the \citet{tremblayetal13-1} 3D corrections. Spectral type classification  with the ``:" suffix are considered uncertain due to the low quality of the spectrum.}
\begin{tabular}{llllllcc}
\hline
name & Ra & Dec & $P_{\mathrm{WD}}$ & instrument & type & \Teff\,(K) & log g\\
\hline
ATLAS\,J034131.17$-$272144.73 & 55.37988 & $-$27.362427 & 0.95 & X-Shooter & DA &$13\,519\pm450$&$7.81\pm0.07$\\
ATLAS\,J001618.70$-$343056.17 & 4.077932 & $-$34.515605 & 0.98 & X-Shooter & DA &$11\,329\pm160$&$7.72\pm0.05$\\
ATLAS\,J000119.76$-$394703.17 & 0.332369 & $-$39.784214 & 0.99 & X-Shooter & DA &$12\,684\pm310$&$8.10\pm0.08$\\
ATLAS\,J000344.78$-$391523.32 & 0.936586 & $-$39.25648 & 0.89 & EFOSC2 &  DA &$6946\pm100$&$7.28\pm0.27$\\
ATLAS\,J002239.01$-$311039.05 & 5.662562 & $-$31.177516 & 1.00 & EFOSC2 &  DAH & &\\
ATLAS\,J002606.30$-$322423.70 & 6.526252 & $-$32.406585 & 0.99 & EFOSC2 &  DA & $11\,708\pm150$&$8.10\pm0.05$\\
ATLAS\,J014005.85$-$344724.11 & 25.02440 & $-$34.790033 & 0.96 & EFOSC2 &  DA &$11\,721\pm630$&$7.76\pm0.28$\\
ATLAS\,J023320.65$-$320310.88 & 38.336047 & $-$32.053023 & 0.99 & EFOSC2 &  DA & $10\,770\pm180$&$7.99\pm0.06$\\
ATLAS\,J023752.56$-$304133.16 & 39.469012 &$-$30.692547 & 0.99 & EFOSC2 &  DAH & &\\
ATLAS\,J034356.22$-$334106.29 & 55.984261 & $-$33.685081 & 0.96 & EFOSC2 &  DA & $12\,419\pm420$&$8.37\pm0.09$\\
ATLAS\,J214039.37$-$341920.25 & 325.164068 & $-$34.322294 & 0.96 & EFOSC2 &  DA &$17\,140\pm230$&$7.85\pm0.06$\\
ATLAS\,J220217.30$-$391728.36 & 330.572104 & $-$39.291212 & 0.98 & EFOSC2 &  DA &$9508\pm90$&$7.94\pm0.08$\\
ATLAS\,J222337.44$-$343839.72 & 335.906028 & $-$34.644369 & 0.87 & EFOSC2 &  DZ & &\\
ATLAS\,J224510.44$-$383645.71 & 341.293532 & $-$38.612699 & 0.88 & EFOSC2 &  DA &$10\,194\pm290$&$7.85\pm0.10$\\
ATLAS\,J224653.56$-$385651.24 & 341.723203 & $-$38.947567 & 	1.00 & EFOSC2 &  DAV &$10\,432\pm290$&$8.06\pm0.10$\\
ATLAS\,J230223.57$-$114811.36 & 345.59821 & $-$11.803158 & 0.98 & EFOSC2 &  DA & $10\,077\pm140$&$7.92\pm0.06$\\
ATLAS\,J034255.41-300122.62 & 55.730916 & $-$30.022952 & 0.99 & 2dF &  DA & $15\,270\pm980$&$9.13\pm0.22$\\
ATLAS\,J033004.84$-$295300.07 & 52.520199 & $-$29.883353 & 0.97 & 2dF &  DA &$17\,880\pm640$ & $8.06\pm0.12$\\
ATLAS\,J034456.50$-$265224.69 & 56.235429 & $-$26.873526 & 0.93 & 2dF &  DA & $35\,860\pm1150$&$8.51\pm0.21$\\
ATLAS\,J034922.82$-$254709.30 & 57.345107 & $-$25.785918 & 0.72 & 2dF &  DA & $7648\pm110$&$7.56\pm0.24$\\
ATLAS\,J035010.83$-$261739.46 & 57.545143 & $-$26.294295 & 0.90 & 2dF &  DA: & &\\
ATLAS\,J121646.04$-$062443.49 & 184.191856 & $-$6.412081 & 0.96 & 2dF &  DA &$7694\pm230$&$7.85\pm0.81$\\
ATLAS\,J121655.61$-$063810.24 & 184.231716 & $-$6.636178 & 0.75 & 2dF &  DA &$9366\pm270$&$8.33\pm0.27$\\
ATLAS\,J121844.60$-$064243.39 & 184.685854 & $-$6.712053 & 0.99 & 2dF &  DA & $19\,750\pm760$&$7.66\pm0.15$\\
ATLAS\,J123540.68$-$074802.08 & 188.919502 & $-$7.800578 & 0.99 & 2dF &  DA & $7950\pm230$&$6.24\pm0.94$\\
ATLAS\,J132001.63$-$074703.50 & 200.00682 & $-$7.784306 & 0.99 & 2dF &  DA & $14\,690\pm1190$&$8.41\pm0.19$\\
ATLAS\,J152811.82$-$145839.45 & 232.049259 & $-$14.977627 & 0.99 & 2dF &  DQ: & &\\
ATLAS\,J234049.50$-$314633.67 & 355.206261 & $-$31.776022 & 0.96 & 2dF &  DA & $9568\pm100$&$7.99\pm0.09$\\
ATLAS\,J234332.65$-$311950.08 & 355.886063 & $-$31.330578 & 0.96 & 2dF &  DA & $13\,240\pm340$&$8.08\pm0.09$\\
ATLAS\,J121912.39$-$071436.07 & 184.801635 & $-$7.243353 & 0.99 & 2dF &  DC: & &\\
ATLAS\,J134211.62$-$073540.1 & 205.548443 & $-$7.594483 & 1.00 & SOAR & DA & $11\,190\pm170$&$8.02\pm0.05$\\
\hline
\end{tabular}
\end{table*}

\section{Conclusion}
We presented the application of our selection method for photometric white dwarfs candidates \citep{gentilefusilloetal15-1} to the latest internal data release of the VST ATLAS survey combined with proper motions from APOP. The resulting catalogue contains \fullcat ATLAS sources with computed $P_{\mathrm{WD}}$. Using a small number of SDSS spectroscopically  confirmed white dwarfs and contaminants  we calculated that  a confidence cut at $P_{\mathrm{WD}} \geq 0.41$ produces a sample of white dwarfs that is 96 per cent complete with an efficiency of 87 per cent. 
We estimate that our catalogue contains $\simeq4200$ high-confidence white dwarf candidates the majority of which have not yet received spectroscopic follow-up. Only $\sim 15$ per cent of the white dwarfs known to date are located in the southern hemisphere and our catalogue therefore constitute a significant improvement on the current North-South knowledge gap. 

Among these thousands of new white dwarfs we expect to find several systems of particular interest: metal polluted white dwarfs (most likely more than 1000 in the final ATLAS footprint) which will improve current statistics on planetary debris abundances, a few tens of white dwarfs with detectable debris discs which can be identified combining our catalogue with IR data from the Vista Hemisphere Survey (VHS, \citealt{VHS13-1}) and WISE \citep{wrightetal10-1}, several magnetic white dwarfs and white dwarfs with rare atmospheric composition (e.g. DQ) like those already identified in our limited spectroscopic follow-up (Sect.\,\ref{spec_follow}), and more pulsating white dwarfs (Sect.\,\ref{pulse}). 
The application of our catalogue to most  white dwarfs population studies will ultimately require spectroscopic follow-up. The possibility to rely on the $P_{\mathrm{WD}}$s allows one to tailor future spectroscopic observations prioritising efficiency (and therefore high $P_{\mathrm{WD}}$ targets) for single target observations or completeness in large scale campaigns.

\section*{Acknowledgements}
{\it Facilities:} VST, AAT, NTT, VLT, SOAR, SDSS
\newline We thank the anonymous referee for the fast and constructive comments received.
\newline The research leading to these results has received funding
from the European Research Council under the European
Union’s Seventh Framework Programme (FP/2007-2013) /
ERC Grant Agreement n. 320964 (WDTracer). 
\newline Support for this work was provided by NASA through Hubble.
Fellowship grant HST-HF2-51357.001-A
\newline Based on observations made with ESO Telescopes at the Paranal Observatory under programme ID 095.D-0406(B), 095.D-0802(B), 097.D-1029(A), 177.A-3011 (A-I).
\newline Based on observations obtained at the Southern Astrophysical Research (SOAR) telescope, which is a joint project of the Minist\'{e}rio da Ci\^{e}ncia, Tecnologia, e Inova\c{c}\~{a}o da Rep\'{u}blica Federativa do Brasil, the U.S. National Optical Astronomy Observatory, the University of North Carolina 
at Chapel Hill, and Michigan State University.
\newline Funding for SDSS-III has been provided by the Alfred P. Sloan Foundation, the Participating Institutions, the National Science Foundation, and the U.S. Department of Energy Office of Science. The SDSS-III web site is http://www.sdss3.org/.
SDSS-III is managed by the Astrophysical Research Consortium for the Participating Institutions of the SDSS-III Collaboration including the University of Arizona, the Brazilian Participation Group, Brookhaven National Laboratory, Carnegie Mellon University, University of Florida, the French Participation Group, the German Participation Group, Harvard University, the Instituto de Astrofisica de Canarias, the Michigan State/Notre Dame/JINA Participation Group, Johns Hopkins University, Lawrence Berkeley National Laboratory, Max Planck Institute for Astrophysics, Max Planck Institute for Extraterrestrial Physics, New Mexico State University, New York University, Ohio State University, Pennsylvania State University, University of Portsmouth, Princeton University, the Spanish Participation Group, University of Tokyo, University of Utah, Vanderbilt University, University of Virginia, University of Washington, and Yale University.

\bibliographystyle{mn_new}

\end{document}